\documentclass[12pt]{article}

\usepackage{scicite}
\usepackage{graphicx}
\usepackage{txfonts}

\usepackage{times}

\newenvironment{sciabstract}{%
\begin{quote} \bf}
{\end{quote}}

\newcounter{lastnote}

\title{Equilibrium analysis of \emph{N}-seller and \emph{N}-buyer Bargaining Games}

\author
{Jiawei Li\\
\\
\normalsize{School of Natural Sciences, University of Stirling, Stirling, UK}\\
\\
\normalsize{E-mail:  lij@cs.stir.ac.uk}
}

\date{}

\begin{document}


\baselineskip24pt


\maketitle


\begin{sciabstract}
  A group of players which contain \emph{n} sellers and \emph{n}
buyers bargain over the partitions of \emph{n} pies. A seller
(/buyer) has to reach an agreement with a buyer (/seller) on the
division of a pie. The players bargain in a system like the stock market: each seller(buyer) can either offer a selling(buying) price to all buyers(sellers) or accept a price offered by another buyer(seller). The offered prices are known to all. Once a player accepts a price offered by another one, the division of a pie between them is determined. Each player has a constant discounting factor and the discounting factors of all players are common knowledge. In this article, we prove that the equilibrium of this bargaining problem is a unanimous division rate for all players, which is equivalent to Nash bargaining equilibrium of a two-player bargaining game in which the discounting factors of two players are the average of \emph{n} buyers and the average of
\emph{n} sellers respectively. This result shows the relevance between bargaining equilibrium and general equilibrium of markets.
\end{sciabstract}


\section*{Introduction}

Game theorists have developed the axiomatic approach and sequential
approach for two-player bargaining games. With a series
of axiomatic assumptions, Nash has proved the unique equilibrium of
two-player bargaining games (Nash 1950). Rubinstein (1982) has developed
the sequential strategic approach in which two players take turns
making alternating offers. In the case where each player has a
constant discounting factor ($\delta_{1}$ and $\delta_{2}$), the
solution is proved to be $(1-\delta_{2})/(1-\delta_{1}\delta_{2})$.
Binmore, Rubinstein and Wolinskyet (1986) discussed the relationship
between these two approaches.

For the bargaining problems with more than two players, the
uniqueness of the perfect equilibrium outcome does not hold even
under the condition of common discounting factor (Sutton 1986;
Herrero 1985; Haller 1986). This bargaining game is described as
multiple players trying to reach an agreement on how to share a pie
between them. Chae and Yang (1988), Yang (1992), Krishna and Serrano
(1996) have proven that the uniqueness of perfect equilibrium can be
achieved by introducing the exit opportunity. Recent related work on
$n\geq3$-person bargaining game also includes Asheim (1992), Winter
(1994), Merlo and Wilson (1995), Dasgupta and Chiu (1998),
Vannetelbosch (1999), Calvo-Armengo (1999), Chatterjee and Sabourian
(2000), Vidal-Puga (2004), De Fontenay and Gans (2004), Kultti
and Vartiainen (2008), Santamaria (2009), Torstensson (2009), and Yan (2009).

In this paper we study a bargaining game with \emph{n}
sellers and \emph{n} buyers. In real world instances, bargaining mostly
occurs between sellers and buyers in exchange. When the market
is not monopolized, each buyer can freely choose among sellers to
make an exchange. However, a buyer(seller) cannot bargain with
another buyer(seller). Therefore, this bargaining game is described
as below. \emph{n} sellers and \emph{n} buyers are about to bargain
on how to share \emph{n} pies. The bargaining process is like the biding system of stock market. Each seller can either offer a selling price to all buyers or accept a buying price offered by any buyer. Similarly, each buyer can either offer a buying price or accept a selling price offered by any seller. Here the price is equivalent to the partition rate of a pie. A pair of seller and buyer will share a pie if they
can reach an agreement. Each player has a constant discounting factor, which means that the value of pie decreases if a player cannot make an agreement with others within a period of time \emph{t}. The discounting factors and offered prices are known to all players.

We prove that this bargaining game has a bargaining
equilibrium with which the bargainers accept a unanimous price
$$p=(n-\sum^{n}_{i=1}\delta_{bi})/(n-\sum^{n}_{i=1}\delta_{si}).$$
where $p$ is the exchange rate or price, $\delta_{si}$ and $\delta_{bi}$ ($i=1,...,n$)
are the discounting factors of the sellers and buyers respectively.

The remainder of this paper is arranged as below. In Section 2, the
preliminary knowledge of sequential approach is introduced. The
advantage of moving first can be eliminated by introducing a bidding
stage in which two players bid for the right to make the first
offer. We also deduce the consistency between the axiomatic approach
and the sequential strategic approach. In Section 3, we prove the bargaining equilibrium of a two-seller and two-buyer bargaining game and extend the solution to emph{n}-seller and emph{n}-buyer case. Section 4 concludes the paper.

\section{Two-player Bargaining Game}

According to Rubinstein (1982), a two-player bargaining game is
described as below:

\emph{Two players, 1 and 2, are bargaining on the
partition of a pie. The pie will be partitioned only after the
players reach an agreement. Each player, in turn offers a partition
and his opponent may agree to the offer or reject it. Acceptance of
the offer ends the bargaining. After rejection, the rejecting player
then has to make a counter offer and so on. If no agreement is
achieved, both players keep their status quo (no gain no
loss).}\vspace{0.2cm}

Let $X$ be the set of possible agreements, $D$ the status quo (no
agreement), and $x_{1}$  and $x_{2}$  the partitions of the pie that
1 and 2 receives respectively. The players' preference relations are
defined on the set of ordered pairs of the type $(x,t)$ , where  $t$ is a nonnegative integer and denotes the time when the agreement is reached, $0 \leq x \leq 1$,  $x_{1}=x$ and $x_{2}=1-x$. Let $\succ_{i}$ denote player \emph{i}'s preference ordering over $X\cup \{D\}$. There are the following assumptions.

\vspace{0.3cm}\noindent\textbf{\emph{A-1}}. Disagreement is the
worst outcome: for every $(x,t)\in X \times T$ we have $(x,t)\succ_{i}D$.\vspace{0.2cm}

\noindent\textbf{\emph{A-2}}. 'Pie' is desirable:
$(x,t)\succ_{i}(y,t)$ iff $x_{i}\geq y_{i}$.\vspace{0.2cm}

\noindent\textbf{\emph{A-3}}. 'Time' is valuable: for every $x\in
X$, $t_{1}<t_{2}$, if $(x,t_{2})\succ_{i}(d,0)$ then
$(x,t_{1})\succ_{i}(x,t_{2})$.\vspace{0.2cm}

\noindent\textbf{\emph{A-4}}. Stationarity: for every $x,y\in X$,
$\Delta >0$, if $(x,t_{1})\succ_{i}(y,t_{1}+\Delta)$ then
$(x,t_{2})\succ_{i}(y,t_{2}+\Delta)$.\vspace{0.2cm}

\noindent\emph{\textbf{A-5}}. Continuity: if
$(x,t_{1})\succ_{i}(y,t_{2})$, there always exists
$\epsilon\rightarrow0$ such that
$(x+\epsilon,t_{1})\succ_{i}(y,t_{2})$.\vspace{0.2cm}

\noindent\textbf{\emph{A-6}}. Increasing loss to delay: for any
$c_{1},c_{2}>0$, if $(x+c_{1},t)\sim_{i}(x,0)$,
$(y+c_{2},t)\sim_{i}(y,0)$ and $x_{i}>y_{i}$ then $c_{1}\geq
c_{2}$.\vspace{0.3cm}

The players are with constant discounting factors: each player has a
number $0\leq\delta_{i}\leq1$ such that
$(x,t_{1})\succ_{i}(y,t_{2})$ iff $x_{i}\delta_{i}^{t_{1}}\geq
y_{i}\delta_{i}^{t_{2}}$. Under these assumptions, Rubinstein (1982)
has proven the following proposition 1.

\vspace{0.3cm}\noindent\emph{\textbf{Proposition 1}}: (a) There
exists a unique perfect equilibrium of this bargaining game (b) If
at least one of the $\delta_{i}$ less than 1 and at least one of
them is positive, the bargaining solution is $(x,0)$, where
$x=(1-\delta_{2})/(1-\delta_{1}\delta_{2})$. \vspace{0.3cm}

Notice that player 1 is supposed to make the first offer. If player
2 makes the first move, the solution would be
$x=(1-\delta_{1})/(1-\delta_{1}\delta_{2})$. The player who makes
the first offer has an advantage in bargaining and receives a larger
partition of the pie than what would be received if another player
had made the first offer.

Binmore, Rubinstein and Wolinsky (1986) gave a procedure to
eliminate the advantage of moving first as follows: let the time
delay between successive periods be $\Delta$, and represent the
discounting factor as $\delta^{\Delta}$. Then in the limit
$\Delta\rightarrow0$, it is indifferent whoever makes the opening
demand.
$$\lim_{\Delta\rightarrow0} x^{*}(\Delta)=\lim_{\Delta\rightarrow0}
y^{*}(\Delta)=x_{N}^{TP}(\succ_{1},\succ_{2})$$

\noindent where $x^{*}(\Delta)$ and $y^{*}(\Delta)$ denote the pair
of agreements and $x_{N}^{TP}(\succ_{1},\succ_{2})$ is the time
preference Nash bargaining solution.

We introduce a new procedure to eliminate the advantage of
moving first as follows: two players bid for the right to make the
first offer before bargaining for the partition of the pie. Player 1
offers a bid $w$ ($0\leq w\leq 1$) to player 2 to exchange the right
of moving first. If player 2 accepts the bid, she receives $w$
partition of the pie and player 1 begins the bargaining to divide
the rest $1-w$; If player 2 refuses the bid, she wins the right to
make the first offer  to divide $1-w$ and player 1 receives $w$. We have the
following proposition 2.

\vspace{0.3cm}\noindent \emph{\textbf{Proposition 2}}: In the case
where two players bid for the right of moving first, the unique
bargaining solution is $(x,0)$, where
$x=(1-\delta_{2})/(2-\delta_{1}-\delta_{2})$.\vspace{0.3cm}

\noindent\textbf{Proof}: If player 2 accepts $w$
and player 1 makes the first move, according to Proposition 1, two
players will receive $x_{1}$ and $x_{2}$ respectively, where
$$x_{1}=\frac{1-\delta_{2}}{1-\delta_{1}\delta_{2}}(1-w)$$
and $x_{2}=1-x_{1}$.

On the other hand, if player 2 declines player
1's bid, two players will receive $x_{1}^{*}$ and $x_{2}^{*}$, where
$$x_{1}^{*}=w+\frac{\delta_{1}(1-\delta_{2})}{1-\delta_{1}\delta_{2}}(1-w)$$
and $x_{2}^{*}=1-x_{1}^{*}$.

It is obvious that there should be $x_{1}=x_{1}^{*}$. Then we have
$$w=\frac{(1-\delta_{1})(1-\delta_{2})}{2-\delta_{1}-\delta_{2}}$$
$$x_{1}=\frac{1-\delta_{2}}{2-\delta_{1}-\delta_{2}}$$
$$x_{2}=\frac{1-\delta_{1}}{2-\delta_{1}-\delta_{2}}$$
\begin{flushright} $\blacksquare$ \end{flushright}

With the bidding procedure, the bargaining equilibrium remains fixed no matter which player makes the first offer and it is indifferent for each player to make an offer or to accept the
opponent's offer. If we regard player 1 as the seller and player 2
the buyer, $p=x_{1}/x_{2}=(1-\delta_{2})/(1-\delta_{1})$ can be
treated as the price of exchange. Specifically, two players share
the pie equally when $\delta_{1}=\delta_{2}$ or $p=1$. Player 2 receives the whole pie when $p=0$. Player
1 receives the whole pie when $p=\infty$. Proposition 2 also shows
the consistency between the strategic bargaining approach and the
Nash bargaining solution. Let $u_{1}$ and $u_{2}$ be the von
Neumann-Morgenstern utilities of two players. If we define
\begin{equation}
p_{N}=\frac{1-\delta_{2}}{1-\delta_{1}}=\mbox{arg max
}(u_{1}(p)u_{2}(p)) \end{equation} the outcome of Proposition 2 is
exactly Nash bargaining solution.

\section{A \emph{N}-seller and \emph{N}-buyer Bargaining Game}

Consider a market where there are only two goods, \emph{A} and
\emph{B}. Participants in this market negotiate to exchange \emph{A}
for \emph{B} or to exchange \emph{B} for \emph{A}. Without loss of
generality, let's assume that those who want to exchange \emph{A}
for \emph{B} are the sellers and those who want to exchange \emph{B}
for \emph{A} are the buyers. The price $p$ is then the exchange
ratio of the amount of \emph{B} to \emph{A} and it is the only
factor that every player cares for in the bargaining. Each seller
prefers a higher price and each buyer prefers a lower price. The
assumptions here are \emph{A-1} to \emph{A-3} and that each player
can freely choose their bargaining partner. The bargaining game is
described as below.

A group of players that contains $n$ sellers and $n$
buyers bargain over the partitions of $n$ pies (each pie is of size
of 1). A seller (/buyer) has to reach an agreement with a buyer
(/seller) on the division of a pie. Each player has a constant
discounting factor and their discounting factors are known to all players($\delta_{si}$ and $\delta_{bi}$ for the sellers
and buyers respectively, $i=1,...,n$, $0<\delta_{si}<1$, $0<\delta_{bi}<1$). The players bargain in a system like the stock market: each seller(buyer) can either offer a selling(buying) price to all buyers(sellers) or accept a price offered by another buyer(seller). The offered prices are known to all. Once a player accepts a price offered by another player, the division of a pie between them is determined and both players quit the
bargaining. The players who do not make any agreement with others remain to the next round and this process will continue until no further agreements are possible or the value of pies decrease to zero.

When an agreement is made, which side offers the price (and which side accepts it) is not important since both sides only care for the price of the agreement. In order to simplify the analysis, we assume that the bargaining process in each round is as follows. The sellers first offer their selling prices. The buyers choose whether or not to accept them. If a buyer does not accept any selling offer, he/she has to offer a buying price. And then the sellers who do not have an agreement choose whether or not to accept a buying price. A round of bargaining ends. The players who do not have an agreement with others enter the next round of bargaining.

We first analyze the bargaining problem of two sellers and two buyers and then extend it to the case of
\emph{n} sellers and \emph{n} buyers.

\subsection{A two-seller and two-buyer bargaining game}\vspace{0.3cm}

Consider the bargaining problem of two sellers $S_{1}$  and  $S_{2}$
and two buyers $B_{1}$  and $B_{2}$ whose discounting factors are
$\delta_{s1}$, $\delta_{s2}$, $\delta_{b1}$ and $\delta_{b2}$
respectively. Let $F$ be the set of all strategies of the players
who offer the partitions, and $G$ the set of all strategies of the
players who have to respond to an offer. The outcome of this
bargaining can be expressed by the quad $(x_{1},t_{1},x_{2},t_{2})$
where $x_{1},x_{2}\in X$ denote the partitions of two pies that two
pairs of players agree with respectively, $t_{1},t_{2}$ denote the time when the agreements are made. Notice that
$t\rightarrow\infty$ denotes 'disagreement' between a seller and a
buyer. We have the following proposition.

\vspace{0.3cm}\noindent \emph{\textbf{Proposition 3}}: A
two-seller and two-buyer bargaining game has a 
bargaining equilibrium, a unanimous partition of both
pies.\vspace{0.3cm}

\noindent \textbf{Proof}: We first prove that, if there exists a bargaining equilibrium, it must be a unanimous partition of two pies (unanimous partition is a necessary condition).

Without loss of generality, suppose that the seller $S_{1}$ and the
buyer $B_{1}$ reach the agreement of partition $(x_{1},t_{1})$ and
$S_{2}$ and $B_{2}$ reach the agreement of partition
$(x_{2},t_{2})$. For the outcome $(x_{1},t_{1},x_{2},t_{2})$ to be a
perfect equilibrium partition, there must be $x_{1}=x_{2}$ and $t_{1}=t_{2}$.

If $x_{1}\neq
x_{2}$, without loss of generality, let's assume $x_{1}>x_{2}$.

(1) If $t_{1}>t_{2}$, $B_{1}$ could offer a buying partition $x'$ ($x_{2}<x'<x_{1}$) at time $t_{2}$ and $S_{2}$  would accept the offer instead of $x_{1}$ so that both players improved their payoffs. Thus, $(x_{1},t_{1},x_{2},t_{2})$ is not an equilibrium.

(2) If $t_{1}< t_{2}$, $S_{2}$ could offer a selling partition $x'$ ($x_{2}<x'<x_{1}$) at time $t_{1}$ and $B_{1}$ would accept it so that both players improved their payoffs. So $(x_{1},t_{1},x_{2},t_{2})$ is not an equilibrium.

(3) If $t_{1}=t_{2}$, either $B_{1}$ or $S_{2}$ would offer a partition $x'$ ($x_{2}<x'<x_{1}$) to improve their payoffs.

Whenever there are $x_{1}\neq x_{2}$, we can always find a pair of seller and buyer who can be better off by making a deal with new partition $x'$ ($x_{2}<x'<x_{1}$). Thus the equilibrium must be a unanimous partition of two pies.

Second, we prove that no player has incentive to deviate from a unanimous partition unilaterally (unanimous partition is a sufficient condition). Given that four players have reached a unanimous partition, in which the partition is $x$ between each pair of seller and buyer. A seller would deviate only if there was a buyer who would accept a higher price than $x$. A buyer would deviate only if there was a seller who would accept a lower price than $x$. Obviously, two conditions are conflict with each other. Thus no player could deviate from a unanimous partition unilaterally.\begin{flushright}$\blacksquare$\end{flushright}

Every player will receive the minimum payoff if they cannot reach a unanimous partition. Realising that they have to reach a unanimous partition, the sellers and buyers actually bargain on a price with which no single player can be better off by deviating from it. Both sellers prefer higher prices to lower prices while both buyers prefer lower prices to higher prices. The bargaining game turns out to be a two-player game, in which one side is the group of sellers and the other side is the group of buyers. According to
Proposition 1, there is a unique equilibrium of the two-player bargaining game. Let $x^*$ denote the partition in this equilibrium. In order to determine $x^{*}$ by adopting a sequential approach, we assume that the sellers first offer a partition to the buyers. If the buyers accept it, two agreements are made and the game ends. Otherwise, the buyers make a counter offer on the discounted pie. The game continues until either two agreements or disagreement is made.

Following Rubinstein (1982) and Osborne and Rubinstein (1990), we
define a group of functions $v_{i}$ $(i=s_{1}, s_{2}, b_{1}, b_{2})$
as follows.
$$v_{i}(x,t)=\left \{ \begin{array}{l l} y, & \exists y \in X \mbox{ such that }
(y,0)\sim_{i}(x,t) \\ 0, & \forall y \in X \mbox{ there is
}(y,0)\succ_{i}(x,t)  \end{array} \right.$$

This means that for any $(x,t)$, either there is a unique $y\in X$
such that player $i$ is indifferent between $(x,t)$ and  $(y,0)$, or
every $(y,0)$ is preferred by $i$ to $(x,t)$. In order for two sellers and two buyers to reach a unanimous partition, there should
be $v_{s_{1}}=v_{s_{2}}$ and $v_{b_{1}}=v_{b_{2}}$. This means that
two sellers (buyers) are equal in the bargaining no matter what
values their discounting factors are.

In order for two sellers (buyers) to form the same bargaining
strategy, there should be
\begin{equation}
v_{s_{1}}(x,t)=v_{s_{2}}(x,t)=\frac{1}{2}(\delta_{s_{1}}^{t}+\delta_{s_{2}}^{t})x
\end{equation}
\begin{equation}
v_{b_{1}}(x,t)=v_{b_{2}}(x,t)=\frac{1}{2}(\delta_{b_{1}}^{t}+\delta_{b_{2}}^{t})x
\end{equation}
The intersection of $y_{s_{1}}=v_{1}(x_{s_{1}},1)$ and
$x_{b_{1}}=v_{1}(y_{b_{1}},1)$ reflects the unanimous partition
$(x^{*},y^{*})$. This can be expressed as Fig.1.

\begin{figure}[hbtp]
  \centering
  \includegraphics[width=240pt]{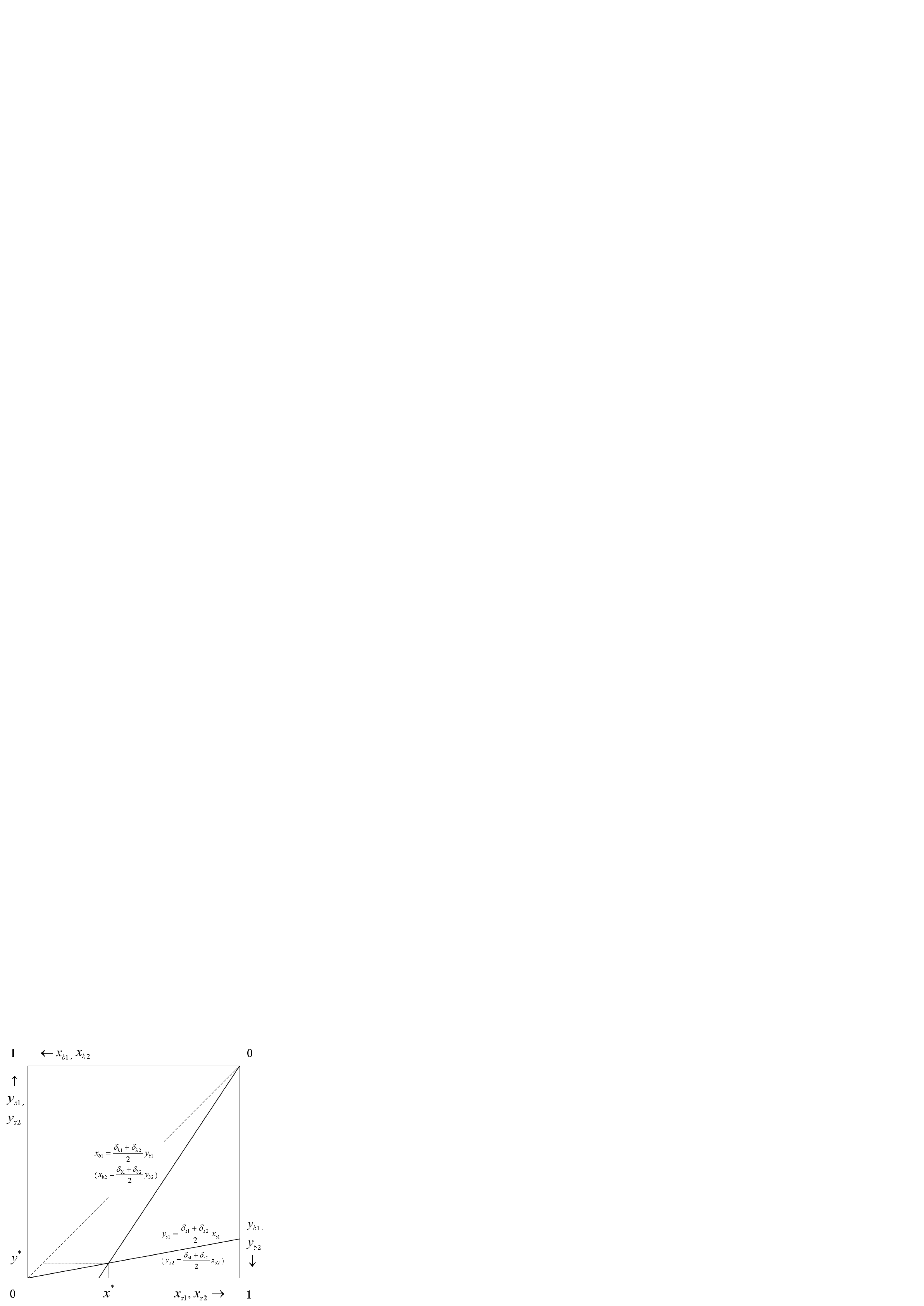}
  \caption{Perfect equilibrium ($x^{*}$, $y^{*}$) for the bargaining game of two-pair of sellers and buyers.}
  \label{Fig1}
\end{figure}

From Fig.1, we have
\begin{equation}
x^{*}=\frac{2(2-\delta_{b1}-\delta_{b2})}{4-(\delta_{b1}+\delta_{b2})(\delta_{s1}+\delta_{s2})}
\end{equation}
\begin{equation}
y^{*}=\frac{(\delta_{b1}+\delta_{b2})(2-\delta_{s1}-\delta_{s2})}{4-(\delta_{b1}+\delta_{b2})(\delta_{s1}+\delta_{s2})}
\end{equation}
When the sellers and buyers bid for the right of making the first
offer, the advantage of first offer can be eliminated. The process
is that the sellers first offers a bid $w$ ($0\leq w\leq 1$) to the
buyers to exchange the right of first offer. If the buyers accept
the bid, each buyer receives $w$ partition of a pie and the sellers
begin the bargaining to divide the rest of pies; If the buyers
refuse the bid, they win the right to make an offer first and each
seller receives $w$.

If the buyers accept the bid, each seller will receive
$$x_{1}=\frac{2(2-\delta_{b1}-\delta_{b2})}{4-(\delta_{s1}+\delta{s2})(\delta_{b1}+\delta_{b2})}(1-w)$$

If the buyers incline the bid, each seller will receive
$$x_{2}=\frac{(\delta_{s1}+\delta_{s2})(2-\delta_{b1}-\delta_{b2})}{4-(\delta_{s1}+\delta{s2})(\delta_{b1}+\delta_{b2})}(1-w)+w$$

Obviously, there should be $x_{1}=x_{2}$. Then we have
\begin{equation}
x_{1}=\frac{2-\delta_{b1}-\delta_{b2}}{4-\delta_{b1}-\delta_{b2}-\delta_{s1}-\delta_{s2}}
\end{equation}
Each buyer receives
\begin{equation}
y_{1}=1-x_{1}=\frac{2-\delta_{s1}-\delta_{s2}}{4-\delta_{b1}-\delta_{b2}-\delta_{s1}-\delta_{s2}}
\end{equation}
Hence,
\begin{equation}
p=x_{1}/y_{1}=\frac{2-\delta_{b1}-\delta_{b2}}{2-\delta_{s1}-\delta_{s2}}
\end{equation}
According to Proposition 3, a patient player receives the same
price as an impatient player in the bargaining. This
counterintuitive-seeming result can be explained as below. Because
every player would like to choose the impatient player to be their
bargaining opponent, the impatient player could increase their share
by threatening to change their bargaining opponent. Similarly, the
patient player had to lower their share because of their opponent's
threat of changing their bargaining opponent. Consequently, a
unanimous price will be reached so that the sellers (buyers) receive
equal partition no matter how patient or impatient they are.

\begin{figure}[hbtp]
  \centering
  \includegraphics[width=240pt,totalheight=220pt]{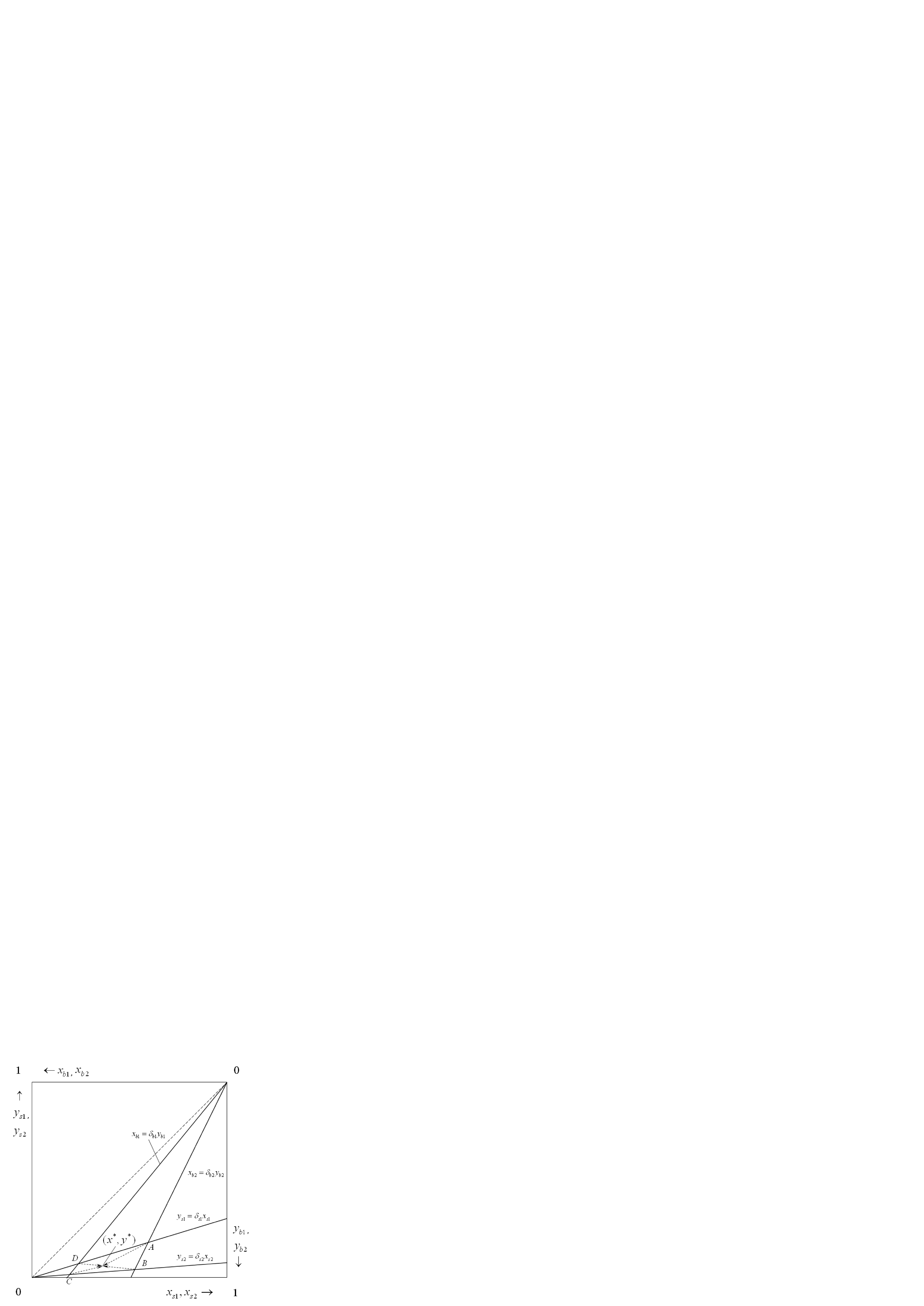}
  \caption{Relationship between two-player bargaining and two-pair of players bargaining. }
  \label{Fig2}
\end{figure}

Fig.2 shows the relationship between the two-player bargaining equilibrium and the two-seller and two-buyer bargaining equilibrium. If the players bargain with each other
independently, the equilibrium of two players bargaining will
be either $A, C$ or $B, D$. When they bargain together, the equilibrium will be $(x^{*},y^{*})$, which is a point inside the quadrilateral $ABCD$.

\subsection{The \emph{n}-seller and \emph{n}-buyer bargaining game}\vspace{0.2cm}

Let $S_{i}$ and $B_{i}$ $(i=1, \cdots,n)$ denote the sellers and
buyers respectively, and $\delta_{s_{i}}$ and $\delta_{b_{i}}$ the
discounting factors of the sellers and buyers respectively. We have
Proposition 4 as below.

\vspace{0.3cm}\noindent \emph{\textbf{Proposition 4}}: The
\emph{n}-seller and \emph{n}-buyer bargaining problem has an
 equilibrium: a unanimous price $p_{n}$,
\begin{equation}
p_{n}=(n-\sum^{n}_{i=1}\delta_{b_{i}})/(n-\sum^{n}_{i=1}\delta_{s_{i}})
\end{equation}

\noindent\textbf{Proof}: According to the proof of Proposition 3, a bargaining equilibrium for this
bargaining game must be a unanimous price. It is easy to verify that, if there are different prices, at least one pair of seller and buyer can be better off by changing their decisions.

Consider the two largest coalitions that contains all sellers and
buyers respectively. Let $p^{*}$ denotes the price determined by
these two coalitions. $p^{*}$ must be an equilibrium because
no player can be better off by leaving their coalition. For
example, a seller would leave the coalition only if there was
a buyer who would accept $p>p^{*}$. However, any buyer would
leave their coalition only if there was $p<p^{*}$. Let $\delta_{s_{A}}$ and
$\delta_{b_{A}}$ be the average discounting factor of the coalitions of
sellers and buyers respectively.
\begin{equation}
\delta_{s_{A}}=(\delta_{s_{1}}+\delta_{s_{2}}+\cdots+\delta_{s_{n}})/n
\end{equation}
\begin{equation}
\delta_{b_{A}}=(\delta_{b_{1}}+\delta_{b_{2}}+\cdots+\delta_{b_{n}})/n
\end{equation}
According to (1), there must be
$$p^{*}=\frac{1-\delta_{b_{A}}}{1-\delta_{s_{A}}}=\frac{n-\sum^{n}_{i=1}\delta_{b_{i}}}{n-\sum^{n}_{i=1}\delta_{s_{i}}}.$$
Then, $p^{*}$ is the unanimous price $p^{*}=p_{n}$.
\begin{flushright}$\blacksquare$\end{flushright}

The equilibrium of the \emph{n}-seller and
\emph{n}-buyer bargaining game can be expressed as Fig.3. Two coalitions reach the agreement
of $(x^{*},y^{*})$. If two coalitions bid for the right to make the
first offer, the advantage of making the first offer will be
removed. Then, each seller receives $x_{1}$,
\begin{equation}
x_{1}=\frac{n-\sum^{n}_{i=1}\delta_{b_{i}}}{2n-\sum^{n}_{i=1}(\delta_{s_{i}}+\delta_{b_{i}})}
\end{equation}
Each buyer receives $y_{1}$,
\begin{equation}
y_{1}=\frac{n-\sum^{n}_{i=1}\delta_{s_{i}}}{2n-\sum^{n}_{i=1}(\delta_{s_{i}}+\delta_{b_{i})}}
\end{equation}
\begin{figure}[hbtp]
  \centering
  \includegraphics[width=240pt,totalheight=220pt]{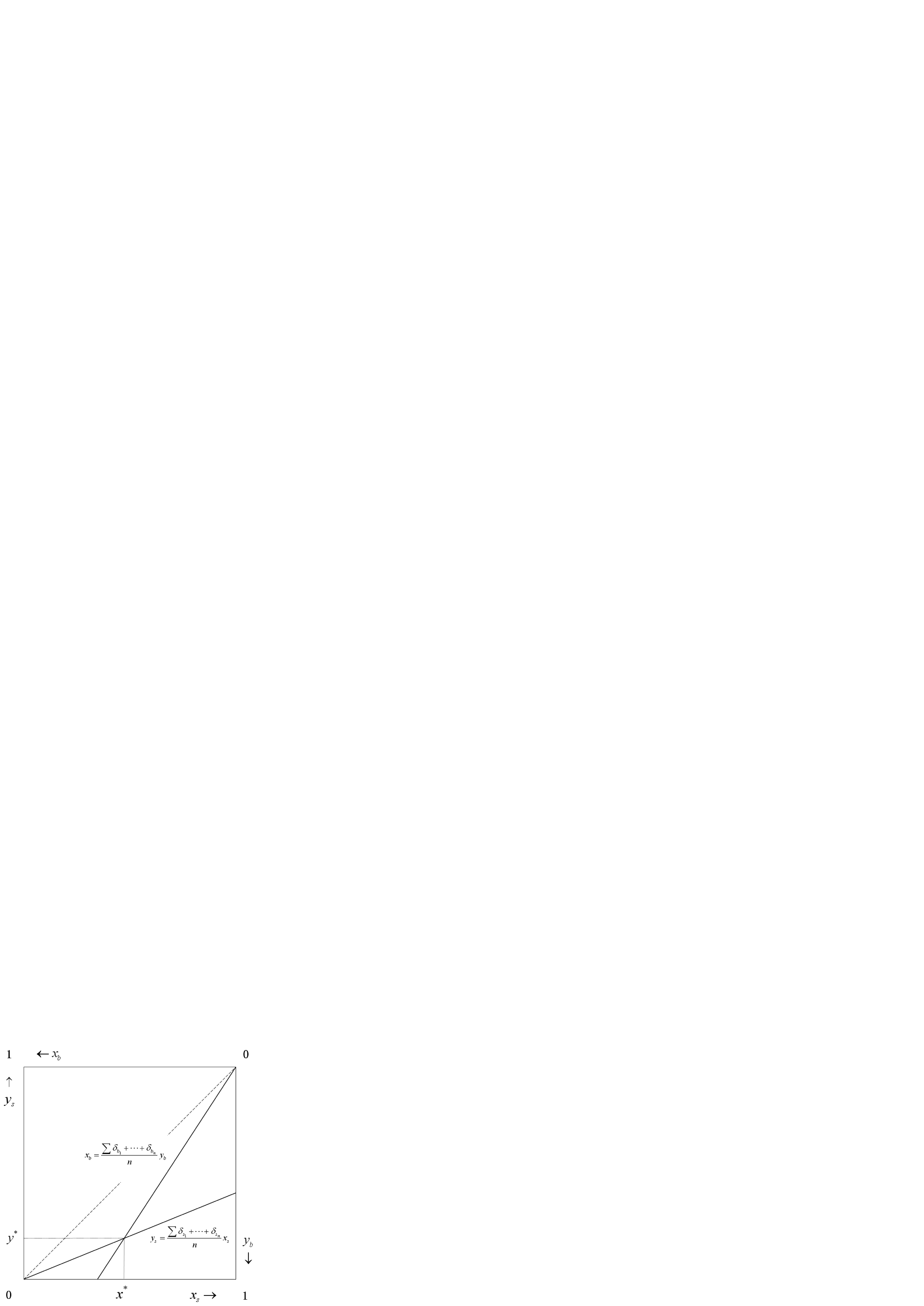}
  \caption{Perfect equilibrium ($x^{*}$, $y^{*}$) for the bargaining game of \emph{n}-pair of sellers and buyers. }
  \label{Fig3}
\end{figure}

Imagine that there exists a pair of representative seller and buyer,
$S_{A}$ and $B_{A}$, whose discounting factors,
$\delta_{s_{A}}$ and $\delta_{b_{A}}$, are the average of discounting factors of all sellers and all buyers respectively. The equilibrium of a
\emph{n}-pair players bargaining game is equivalent to the equilibrium of
two representative players bargaining game. Let $u_{s_{A}}(p)$ and
$u_{b_{A}}(p)$ be the von Neumann-Morgenstern utility functions of
the representative seller and buyer respectively. If we define
$$p_{N}=\frac{n-\sum^{n}_{i=1}\delta_{b_{i}}}{n-\sum^{n}_{i=1}\delta_{s_{i}}}=\mbox{arg max }(u_{s_{A}}(p)u_{b_{A}}(p)),$$
the outcome of Proposition 4 is consistent with Nash bargaining
equilibrium.

\section{Conclusions}

The bargaining games of two-seller and two-buyer, and further \emph{n}-seller and \emph{n}-buyer are analyzed by using a sequential approach. The equilibria of these
bargaining games are unanimous prices determined by average discounting factors of all sellers and all buyers. Every player is a price taker when \emph{n} is large because individual player has trivial power to determine
the bargaining price.
  This conclusion has the potential of being extended to the
problems of market-clearing prices in perfectly competitive markets.

In this study, the $2n$ players are assumed to bargain together so
that the bargaining problem can be considered as a game with complete
information in which each player knows the discounting factors of all players. This may be unrealistic in real world
market, especially when the number of players is relatively
large. It sometimes takes time, money, and other resources for the players to
retrieve certain information. If the cost of information is taken
into consideration, the players will have to restrict their negotiations
within a limited group of players. Then the propagation of
information will have an influence on the players' strategies and
different prices are possible in this circumstance.

\bibliography{scibib}
\bibliographystyle{Science}

\vspace{1.0cm}\noindent {\normalfont{\textbf{References}}}

\noindent 1. Asheim, G. (1992) A unique solution to n-person sequential
bargaining, Games and Economic Behavior, 4: 169-181.

\noindent 2. Binmore, K., Rubinstein A., and Wolinsky, A. (1986) The Nash
Bargaining Solution in Economic Modelling, RAND Journal of
Economics, 17:176-188.

\noindent 3. Calvo-Armengo, A. (1999) A note on three-player noncooperative
bargaining with restricted pairwise meetings, Economics Letters, 65:
47-54.

\noindent 4. Chae, S. and Yang, J. (1988) The unique perfect equilibrium of an N-person
bargaining game, Economics Letters, 28: 221-223.

\noindent 5. Chatterjee, K. and Sabourian H. (2000) Multiperson bargaining and
strategic complexity, Econometrica, 68: 1491-1509.

\noindent 6. Dasgupta, A. and Chiu, Y. (1998) On
implementation via demand commitment games, International Journal of
Game Theory, 27: 161-189.

\noindent 7. de Fontenay, C. and Gans J. (2004) Bilateral Bargaining with
Externalities, Working Paper (University of Melbourne).

\noindent 8. Haller, H. (1986) Non-cooperative bargaining of NU3 players,
Economics Letters, 22: 11-13.

\noindent 9. Herrero, M. (1985) A strategic bargaining approach to market
institutions, Ph.D. thesis (London University, London).

\noindent 10. Krishna, V. and Serrano, R. (1996) Multilateral bargaining, Review of
Economic Studies, 63: 61-80.

\noindent 11. Kultti K. and Vartiainen H. (2008) Bargaining with many players: a
limit result, Economics Letters, 101: 249-252.

\noindent 12. Merlo, A. and Wilson, C. (1995) A stochastic model of sequential
bargaining with complete information, Econometrica, 63: 371-399.

\noindent 13. Nash, J. (1950) The Bargaining Problem,
Econometrica, 18: 155-162.

\noindent 14. Osborne, M. and Rubinstein, A. (1990) Bargaining and markets,
Academic Press, California, ISBN 0-12-528632-5.

\noindent 15. Rubinstein, A. (1982) Perfect equilibrium in a bargaining model,
Econometrica, 50(1): 97-109.

\noindent 16. Santamaria, J. (2009) Bargaining power in the Nash damand game an evolutionary approach, International game theory review, 11(1): 87-97.

\noindent 17. Sutton, J. (1986) Non-cooperative bargaining theory: an
introduction, Review of Economic Studies, 53: 709-724.

\noindent 18. Torstensson, P. (2009) An n-person Rubinstein bargaining game, International game theory review, 11(1): 111-115.

\noindent 19. Vannetelbosch, V. (1999) Rationalizability and equilibrium in
n-person sequential bargaining, Economic Theory, 14: 353-371.

\noindent 20. Vidal-Puga, J. (2004) Bargaining with commitments, International
Journal of Game Theory, 33: 129-144.

\noindent 21. Winter, E. (1994) The demand commitment bargaining and snowballing
cooperation, Economic Theory, 4: 255-273.

\noindent 22. Yan, H. (2009) Uniqueness in random-proposer multilateral bargaining, International game theory review, 11(4): 407-417.

\noindent 23. Yang, J. (1992) Another N-person bargaining game with a unique perfect
equilibrium, Economics Letters, 38: 275-277.



\end{document}